\documentclass[%
 reprint,
superscriptaddress,
 amsmath,amssymb,
 aps,
 prl,
]{revtex4-2}

\usepackage{lipsum,xcolor,units}
\usepackage{graphicx}% Include figure files
\usepackage{dcolumn}% Align table columns on decimal point
\usepackage{bm}% bold math
\usepackage{hyperref}% add hypertext capabilities
\usepackage{float}
\begin{document}

\preprint{APS/123-QED}

\title{Precision constraints on the neutron star equation of state with\\
third-generation gravitational-wave observatories}

\author{Kris Walker}
\email{kris.walker@icrar.org}
\affiliation{International Centre for Radio Astronomy Research, University of Western Australia, 35 Stirling Highway, Crawley WA 6009, Australia}
\affiliation{OzGrav: The ARC Centre of Excellence for Gravitational Wave Discovery, Clayton VIC 3800, Australia}

\author{Rory Smith}
\email{rory.smith@monash.edu}
\affiliation{OzGrav: The ARC Centre of Excellence for Gravitational Wave Discovery, Clayton VIC 3800, Australia}
\affiliation{School of Physics and Astronomy, Monash University, VIC 3800, Australia}

\author{Eric Thrane}
\affiliation{OzGrav: The ARC Centre of Excellence for Gravitational Wave Discovery, Clayton VIC 3800, Australia}
\affiliation{School of Physics and Astronomy, Monash University, VIC 3800, Australia}

\author{Daniel J. Reardon}
\affiliation{OzGrav: The ARC Centre of Excellence for Gravitational Wave Discovery, Clayton VIC 3800, Australia}
\affiliation{Swinburne University of Technology, PO Box 218, Hawthorn VIC 3122, Australia}


\date{\today}

\begin{abstract}
It is currently unknown how matter behaves at the extreme densities found within the cores of neutron stars.
Measurements of the neutron star equation of state probe nuclear physics that is otherwise inaccessible in a laboratory setting.
Gravitational waves from binary neutron star mergers encode details about this physics, allowing the equation of state to be inferred.
Planned third-generation gravitational-wave observatories, having vastly improved sensitivity, are expected to provide tight constraints on the neutron star equation of state. 
We combine simulated observations of binary neutron star mergers by the third-generation observatories Cosmic Explorer and Einstein Telescope to determine future constraints on the equation of state across a plausible neutron star mass range. 
In one year of operation, a network consisting of one Cosmic Explorer and the Einstein Telescope is expected to detect $\gtrsim 3\times10^5$ binary neutron star mergers.
By considering only the 75 loudest events, we show that such a network will be able to constrain the neutron star radius to at least $\lesssim\unit[200]{m}$ ($90\%$ credibility) in the mass range $\unit[1-1.97]{M_{\odot}}$---about ten times better than current constraints from LIGO-Virgo-KAGRA and NICER.
The constraint is $\lesssim\unit[75]{m}$ ($90\%$ credibility) near $\unit[1.4-1.6]{M_{\odot}}$ where we assume the the binary neutron star mass distribution is peaked. 
This constraint is driven primarily from the loudest $\sim$20 events.
\end{abstract}

\maketitle

\textit{Introduction.}---The cores of neutron stars host the densest baryonic matter in the Universe.
Travelling from the neutron star surface down toward the core, matter first forms a homogeneous neutron liquid before the appearance of strange baryons and/or deconfined quarks; see \cite{Burgio} for a review.
Theoretical calculations of nuclear physics describing the interiors of neutron stars is notoriously difficult, and laboratory experiments do not begin to approach the necessary densities.
Astronomical measurements of the neutron star equation of state therefore provide a unique probe of nuclear physics at the most extreme possible densities.

Gravitational waves contain rich information about the tidal deformations experienced by coalescing neutron stars, encoding details about the mysterious physics of their interiors \cite{Yu2023}. The susceptibility of a neutron star to deformation by tidal forces is determined by the equation of state of the material composing the star and is quantified by the dimensionless tidal deformability
\begin{align}
    \Lambda=\frac{2}{3}\,k_2\left(\frac{c^2 R}{Gm}\right)^5 .
\end{align}
Here, $k_2$ is the tidal Love number while $m$ and $R$ are the mass and radius of the star. 
The effects of deformations on the gravitational waveform are subtle. Nevertheless, current observations have ruled out some of the ``stiffest'' proposed equations of state \cite{Abbott2018,Radice2018}\footnote{A ``stiffer'' equation of state implies that the neutron star is more resistant to contraction under the influence of its own gravity, and so stiffer equations of state tend to produce relatively larger neutron stars than soft equations of state.}. 
These constraints are expected to improve dramatically with the advent of third generation observatories such as Cosmic Explorer \cite{Evans,Abbott2017,Reitze2019} and the Einstein Telescope \cite{Punturo2010,Hild2011,Maggiore2020}, which aim to probe gravitational-wave strains more than an order of magnitude weaker than possible with current observatories. 
However, quantifying the expected improvement across the full range of binary neutron star masses has proven to be a challenge owing to the computational difficulties associated with analysis of long, high signal-to-noise-ratio gravitational-wave signals \cite{Bayes3G}.

The measurability of the neutron star equation of state has been explored in a number of works in the context of current observatories \cite{Markakis2012,DelPozzo2013,Wade2014,Lackey2015,Agathos2015,Hotokezaka2016,Kumar2017,HernandezVivanco2019}.
In the first fully Bayesian analysis, \cite{DelPozzo2013} considered a linear approximation of $\Lambda(m)$ and found that a few tens of mock events observed by the Advanced LIGO-Virgo network are sufficient to constrain the tidal deformability to 10\% accuracy at a reference mass of $1.4\,{\rm M}_{\odot}$. 
This work was extended in \cite{Lackey2015}, which considered a piecewise polytropic parametrization of the equation of state and showed that---with a network of two Advanced LIGO and one Advanced VIRGO detectors---$\Lambda$ can be constrained to 10\%-50\% accuracy across the mass range $\unit[1-2]{M_\odot}$.
Reference~\cite{HernandezVivanco2019} added additional realism, estimating the constraints available after the first 40 binary neutron star detections.

Some recent work has begun to establish the constraints that will be possible with third-generation observatories. Ref. \cite{Pacilio2022} illustrates the ability of Cosmic Explorer or the Einstein Telescope to discriminate between equation of state models. Ref. \cite{Gupta2022} predicts constraints on the pressure by the Einstein Telescope, while \cite{Ghosh2022,Finstad2023,Pradhan2023,Pradhan2023b} predict the constraint on the radius with third-generation observatories.

In this \textit{Letter}, we determine the ability of an array of third-generation gravitational-wave observatories to constrain the neutron star equation of state. 
We simulate the loudest binary neutron star events from one year of coincident data from the third-generation observatories Cosmic Explorer and the Einstein Telescope.
For each event, we obtain a measurements of the mass and tidal deformability. 
The dependence of the tidal deformability on the mass, $\Lambda(m)$, is uniquely determined by the equation of state through the Tolman–Oppenheimer–Volkoff equation.
Using a spectral decomposition, we apply hierarchical inference to constrain the equation of state across the mass range $\unit[1-1.97]{M_\odot}$.

\textit{Method.}---Our first step is to generate a set of the 75 loudest binary neutron star mergers likely to be observed by a observatory array consisting of one Cosmic Explorer and one Einstein Telescope in one year of coincident data \cite{Abbott2017}.
The merger distances are sampled from the neutron star merger rate density model given in \cite{Safarzadeh2019}, which is parameterized by the minimum merger timescale $t_{\rm min}$ and exponent $\alpha$ of the merger time distribution ${\rm d}N/{\rm d}t_{\rm merger}=t^\alpha$. While there is significant freedom in the choice of these parameters, the resulting difference in detection rate is insignificant in the case of Cosmic Explorer + Einstein Telescope at the low redshifts of the 75 loudest events. We therefore simply choose $t_{\rm min}=\unit[10]{Myr}$ and $\alpha=-1/2$.
The other extrinsic parameters are drawn from standard distributions.
In light of the low observed spins of neutron star binaries \cite{Burgay2003,Stovall2018,Abbott2019}, we take all the neutron stars to have zero spin for simplicity (though we note that the presence of spin can have an effect on the inferred equation of state \cite{Harry2018,Wysocki2020}).
We sample the masses from a Gaussian approximation to the observed Galactic neutron star mass distribution \cite{Kiziltan2013}, though, see \cite{bns-mass}. For this analysis we assume an \texttt{SLy} equation of state \cite{Douchin2001}, which determines the tidal deformabilities from the masses. 
For each event, we calculate the gravitational waveform using the \texttt{IMRPhenomPv2\_NRTidal} approximant included in the LALSuite software suite \cite{lalsuite}.

\begin{figure}
    \centering
    \includegraphics[scale=0.9]{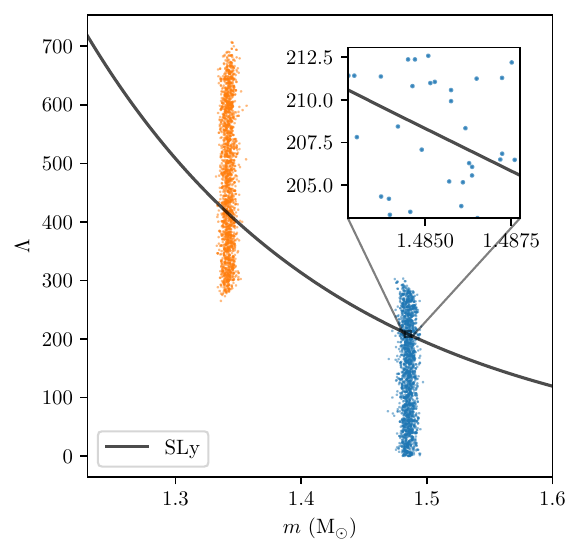}
    \caption{The marginalized $m$-$\Lambda$ posteriors (blue and orange points) from parameter estimation on a merger event overlayed on the black $\Lambda(m)$ curve predicted by the \texttt{SLy} equation of state. Because individual measurements produce a set of (zero-dimensional) discrete posterior samples, there is vanishingly small probability that any equation of state curve will pass through even a single posterior sample (see inset). We therefore estimate a continuous probability density from the samples that can be integrated along a curve passing through it.}
    \label{fig:mass_lambda_posterior}
\end{figure}

To measure the equation of state from the mock events, we use the following procedure:
\begin{enumerate}
    \item We use the \texttt{dynesty} dynamic nested sampling package \cite{Higson2018} included in \texttt{Bilby} \cite{bilby,bilby_gwtc1,Smith_2020_pbilby} to perform Bayesian inference on each event $i$ with data $d_i$. We use the \texttt{ROQGravitationalWaveTransient} likelihood, which implements a reduced-order quadrature integration rule to greatly speed up evaluation \cite{Smith2016,Bayes3G}.
    This yields posterior distribution samples for the binary parameters, $\theta_i$. 
    For this investigation, we are interested only in the component masses $m_1, m_2$ and the tidal deformabilities $\Lambda_1, \Lambda_2$. 
    We therefore marginalize over the other parameters to obtain the marginal posteriors, which are proportional to the marginal likelihoods ${\cal L}(d_i | m_1^i, m_2^i, \Lambda_1^i, \Lambda_2^i)$. An example is shown in Figure \ref{fig:mass_lambda_posterior}.
    %%%%
    \item As discussed in~\cite{HernandezVivanco2019}, it is necessary to interpolate between posterior samples in order to integrate the ${\cal L}(d_i | m_1^i, m_2^i, \Lambda_1^i, \Lambda_2^i)$ along the different equation-of-state curves.
    We therefore use a kernel density estimate (KDE)---calculated using an Epanechnikov kernel---to obtain a continuous representation of the marginal likelihood function~\footnote{Experimentation with other kernels suggests that our results are robust to the choice of kernel}.
    \item To obtain the marginal likelihood for the data given a particular equation of state, we integrate each ${\cal L}(d_i | m_1^i, m_2^i, \Lambda_1^i, \Lambda_2^i)$ along the predicted $\Lambda(m)$ curve and take their product to obtain the total likelihood.
\end{enumerate}

% \begin{figure}
%     \centering
%     \includegraphics[scale=0.6]{figures/kde_accuracy_event_10.pdf}
%     \caption{Comparison between the distribution of the marginalized posterior samples (blue histogram) and their kernel density estimate (orange curve), for one of the events in our sample.}
%     \label{fig:kde_comparison}
% \end{figure}

To model the equation of state, we use a four-parameter spectral decomposition. In this representation, the adiabatic index is given as a function of the pressure $p$ by
\begin{align}
\Gamma(p)=\exp\left(\sum_{k=0}^3\gamma_k \ln(p/p_0)^k\right),
\end{align}
where $p_0$ is a reference pressure that we take to be $\unit[1.64\times 10^{32}]{Pa}$ and $\gamma_k$ are coefficients determined by the equation of state. 
The equation of state (energy density as a function of pressure) $\epsilon(p)$ is obtained by solving
\begin{align}
    \frac{\epsilon+p}{p}\frac{{\rm d}p}{{\rm d}\epsilon}=\Gamma(p).
\end{align}
Truncating the spectral decomposition at four terms has been shown to produce reasonably good fits to realistic equations of state, including \texttt{SLy} \cite{Lindblom2010}.

As mentioned above, the equation of state---and hence the set of parameters $\Upsilon=\{\gamma_0,\gamma_1,\gamma_2,\gamma_3\}$---determines how the tidal deformability depends on mass: $\Lambda(\Upsilon;m)$. 
The likelihood for the full set of data $d$ given these parameters is
\begin{align}
    \mathcal{L}(d\,|\,\Upsilon)=&\prod_{i=1}^{N}\int\mathrm{d}m_1^i
    \int \mathrm{d}m_2^i\,\pi(m_1^i,m_2^i)\nonumber\\
    &\mathcal{L}_\kappa\Big(d_i\,\Big|\,m_1^i, m_2^i, \Lambda(\Upsilon;m_1^i), \Lambda(\Upsilon;m_2^i)\Big),\label{eq:likelihood1}
\end{align}
where $\mathcal{L}_\kappa(d_i\,|\,\cdots)$ are the single-event likelihoods.
Meanwhile, $\pi(m_1^i,m_2^i)$ is the prior on the component masses, which we take to be the distribution from \cite{Kiziltan2013} used to simulate our population of binary neutron stars. 
We evaluate this marginal likelihood with a Riemann sum over bins $a,b$ of each KDE $\mathcal{K}_i$:
\begin{align}
    \mathcal{L}(d&\,|\,\Upsilon)\approx\prod_{i=1}^{N}\Delta m_1^i\Delta m_2^i\sum_{a,b}\,\pi(m_1^{a},m_2^{b}) \nonumber\\
    & \mathcal{K}_i(m_1^{a}, m_2^{b}, \Lambda(\Upsilon;m_1^{a}), \Lambda(\Upsilon;m_2^{b})).
\end{align}

\begin{figure}[H]
    \centering
    \includegraphics[scale=0.9]{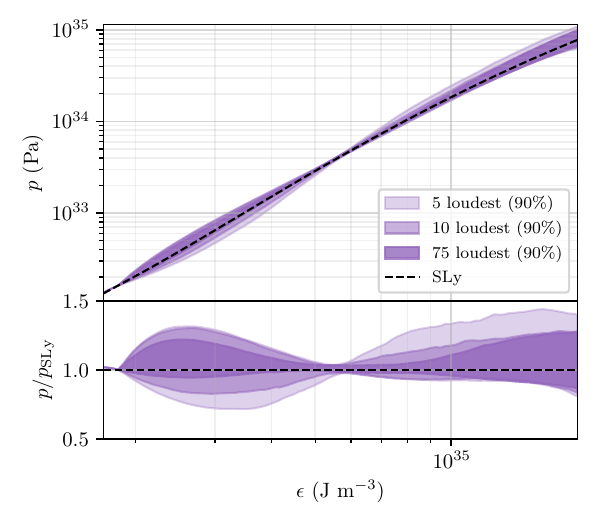}
    \includegraphics[scale=0.9]{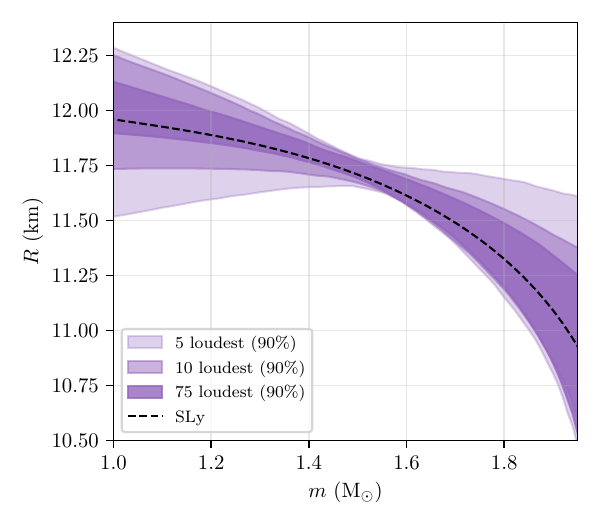}
    \caption{
    Constraints on  the neutron star equation of state.
    The shaded purple regions show the $90\%$ credible intervals.
    The light contours are derived using only the 5 loudest events; the medium contours using the 10 loudest events; and the dark contours using the 75 loudest events.
    Top: pressure $p$ as a function of energy density $\epsilon$.
    The upper panel shows the pressure in units of Pa. 
    The lower panel shows the pressure relative to the \texttt{SLy} equation of state used to generate the data.
    Bottom: the radius $R$ as a function of mass $m$.
    In all three panels, the \texttt{SLy} curve is indicated with the dashed black curve.
    }
    \label{fig:pressure_radius}
\end{figure}

\textit{Results \& Discussion.}---The $90\%$ credible intervals for $p(\epsilon)$ and $R(m)$ are shown in Fig.~\ref{fig:pressure_radius}.
The different shading shows how the constraints vary depending on the number of events used in the fit: the 5 loudest (light), the 10 loudest (medium), and the 75 loudest (dark).
The \texttt{SLy} equation of state used to generate the data is the dashed black curve. 
The dashed black curve is enclosed within the one-sigma credible interval, indicating that we successfully estimate the true equation of state.

The 75 loudest events allow us to constrain $R(m)$ to within an average of $\sim\unit[200]{m}$ over the interval $\unit[1-1.97]{M_\odot}$. 
The constraint is $\sim\unit[75]{m}$ around $\unit[1.4-1.6]{M_\odot}$, near where our distribution of binary neutron stars peaks. 
We constrain the pressure to within an average of $\sim 18\%$ in the energy density range $\unit[2\times 10^{34}\, -2\times 10^{35}]{J\, m^{-3}}$.
The constraint is $\sim 4\%$ at $\epsilon\approx \unit[5.2\times 10^{34}]{J\, m^{-3}}$.

Our ability to constrain the equation of state starts to plateau at around the first $20$ loudest events.
Including additional events improves the constraints, but with diminishing returns.

\begin{figure}[H]
    \centering
    \includegraphics[scale=0.9]{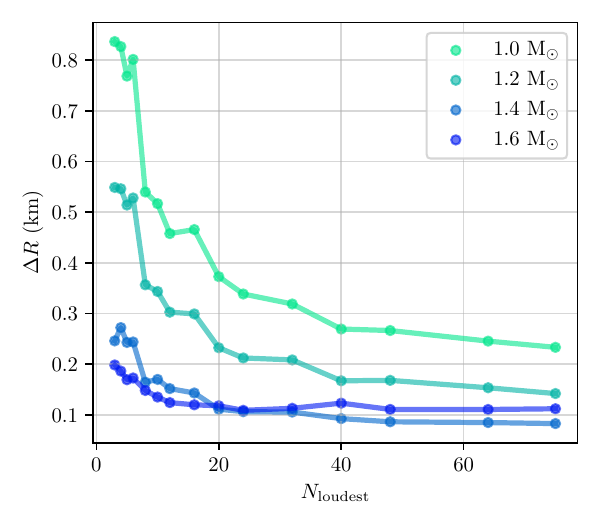}
    \caption{The width of the $90\%$ $R(m)$ credible interval as a function of the number of loudest events used in the analysis $N_\text{loudest}$.
    The improvement in the credible interval begins to plateau at around $N_\text{loudest}=20$.
    }
    \label{fig:constraint_vs_nloudest}
\end{figure}

Figure \ref{fig:constraint_vs_nloudest} shows the width of the $R(m)$ credible interval as a function the number of loudest events included in the analysis $N_\text{loudest}$. 
The width shows little change beyond $N_\text{loudest} \gtrsim 20$ for all values of the mass. 
We conclude that the loudest events play an outsize role constraining the equation of state.
However, given the sheer volume binary neutron star detections ($\approx 3\times 10^5$ per year of Cosmic Explorer \cite{Evans}), the effect of so many small improvements may become significant.
Indeed, the curves are well-fit by both a decaying exponential and a power-law with exponent $\sim -1/2$. The former predicts negligible improvement even with the full set of data, while the latter predicts an improvement by at least a factor of two.
Our sensitivity estimates are therefore conservative.
A mock study to estimate the sensitivity gained from including every binary neutron star detected by Cosmic Explorer and the Einstein Telescope would require significant computational resources.

These results are based on the \texttt{SLy} equation of state and so are representative of constraints achievable for smooth equations of state that are well-fit by parametric models. However, the true equation of state may not be so ``nice'' and could potentially exhibit discontinuous behaviour in the speed of sound due to e.g. phase transitions \cite{Landry2019,Raithel2023}. This can result in uncertainties in derived quantities such as the radius being underreported when using smooth parametric models \cite{Legred2022,Legred2023}. Next-generation X-ray pulse profile observations---such as those by the planned STROBE-X---will allow neutron star radii to be measured to $\sim 2-4\%$ \cite{Ray2019}, providing a valuable check on the constraints from tidal deformability measurements using gravitational waves.

\textit{Comparison with current constraints.}---Observations of the binary neutron star merger GW170817 by LIGO-Virgo \cite{GW170817} constrained the neutron-star radius with a precision of $\unit[2.8]{km}$ at 90\% credibility  \cite{Abbott2018}.
Our results therefore suggest that a network consisting of Cosmic Explorer + Einstein Telescope will improve on this constraint by a factor of $\gtrsim 10$ after 1 year of observations. 
The neutron star equation of state is also constrained by the Neutron Star Interior Composition Explorer (NICER) and X-ray Multi-Mirror (XMM-Newton) observatories, using fits to emission from rotating hot spots \cite{Miller2019,Miller2021}. 
When the NICER and XMM-Newton measurements are combined with the tidal deformability constraints from GW170817 \cite{Abbott2018} and GW190425 \cite{Abbott2020}, the radius at $1.4\,\,{\rm M}_{\odot}$ is constrained to  to 16\% at 90\% credibility \cite{Miller2021,Raaijmakers2021,Pang2021}
The constraint from third-generation gravitational-wave observatories will be $\lesssim 2\%$.
Of course, in this work we include only the 75 loudest events detected in one year of data.
The constraints obtained from the addition of hundreds of thousands of weaker events will improve the constraints by an unknown amount.

\textit{Limitations}---There are some limitations with this analysis that are worthy of further investigation.

The binary neutron star merger rate remains highly uncertain \cite{Kim2015,GW170817}, and it is unclear how changing the merger rate model---and hence the distribution of distances---will affect the equation of state constraints.

For this analysis we use a Gaussian approximation to the observed Galactic binary neutron star distribution. However, this is unlikely to be representative of the extragalactic distribution, with measurements of GW190425 \cite{Abbott2020} suggesting a that it is broader than the Galactic distribution \cite{Landry2021}. It is unclear how this difference will affect the constraint on the equation of state.

Finally, we note that these results are only as accurate as the waveforms used to produce the mock measurements. Such precise measurements require highly accurate waveforms to avoid mismodelling biases, potentially making the effects of additional resonant modes and higher post-Newtonian orders significant \cite{Owen2023,Hu2022}. Recently, \cite{Gupta2022} has shown that including the effects of resonant r-modes in Einstein Telescope observations has a noticeable impact on the inferred equation of state.

\begin{acknowledgments}
We thank Paul Lasky, Aditya Vijaykumar, Ling Sun, Isaac Legred, and Tathagata Ghosh for helpful comments.
This work is supported through Australian Research Council (ARC)  Centre of Excellence CE170100004, Discovery Project and DP230103088, and LIEF Project LE210100002.
This work was performed on the OzSTAR national facility at Swinburne University of Technology. The OzSTAR program receives funding in part from the Astronomy National Collaborative Research Infrastructure Strategy (NCRIS) allocation provided by the Australian Government, and from the Victorian Higher Education State Investment Fund (VHESIF) provided by the Victorian Government.
\end{acknowledgments}

\bibliography{references}

\end{document}